\documentclass[11pt,amsmath,amssymb,nofootinbib,aps,prl]{revtex4}

\usepackage{graphicx}

\begin{document}

\title{OPERA superluminal neutrinos and Kinematics in Finsler spacetime}

\author{Zhe Chang}
\email{changz@ihep.ac.cn}
\author{Xin Li}
\email{lixin@ihep.ac.cn}
\author{Sai Wang}
\email{wangsai@ihep.ac.cn}

\affiliation{Institute of High Energy Physics\\
Theoretical Physics Center for Science Facilities\\
Chinese Academy of Sciences\\
100049 Beijing, China}


\begin{abstract}
The OPERA collaboration recently reported that muon neutrinos could be superluminal.
More recently, Cohen and Glashow pointed that such superluminal neutrinos would be suppressed
since they lose their energies rapidly via bremsstrahlung.
In this Letter, we propose that Finslerian nature of spacetime could account for the superluminal phenomena of particles.
The Finsler spacetime permits the existence of superluminal behavior of particles while the casuality still holds.
A new dispersion relation is obtained in a class of Finsler spacetime.
It is shown that the superluminal speed is linearly dependent on the energy per unit mass of the particle. 
We find that such a superluminal speed formula is consistent with data of OPERA, MINOS and Fermilab-1979 neutrino experiments
as well as observations on neutrinos from SN1987a.

\end{abstract}


\maketitle


Recently, the OPERA Collaboration \cite{OPERA01} reported some evidences that muon neutrinos could propagate superluminally.
The speed difference between neutrinos and light is reported to be
\begin{equation}
\label{OPERA data}
\delta v:=v-1=\left(2.48\pm0.28(\rm{stat.})\pm0.30(\rm{sys.})\right)\times10^{-5}\
\end{equation}
with a significance of \(6\sigma\). Here, we used the natural units as \(c=1\).
The data could be split into two groups and the superluminal speed difference could be given by
\begin{eqnarray}
\label{OPERA splitting data}
\delta v=
\begin{cases}
\left(2.16\pm0.76(\rm{stat.})\pm0.30(\rm{sys.})\right)\times10^{-5}\ ,  ~~~\rm{for~ \langle E\rangle=13.9~GeV}\ ;\\
\left(2.74\pm0.74(\rm{stat.})\pm0.30(\rm{sys.})\right)\times10^{-5}\ ,  ~~~\rm{for~ \langle E\rangle=42.9~GeV}\ .
\end{cases}
\end{eqnarray}
Some earlier experiments gave similar reports on the superluminal phenomena of neutrinos.
However, these earlier experiments have lower significance than OPERA experiment.
The MINOS experiment \cite{MINOS01} in Fermilab claimed that the superluminal speed difference was
\begin{eqnarray}
\label{MINOS data}
\delta v=\left(5.1\pm 2.9(\rm{stat.+sys.})\right)\times10^{-5}\ ~(68\%~\rm{C. L.})\ ,\\
-2.4\times10^{-5}<\delta v<12.6\times10^{-5}\ ~(99\%~\rm{C. L.})
\end{eqnarray}
for muon neutrinos with energies around \(3~\rm{GeV}\) in \(2007\).
Claims from the earlier Fermilab experiment in 1979 (Fermilab1979) show that
the superluminal speed difference is at the similar order of \(10^{-5}\)
for muon neutrinos with energies from \(30~\rm{GeV}\) to \(120~\rm{GeV}\) \cite{Fermilab197901,Fermilab197902}.
In addition, the observations of electron neutrinos from Supernovae-1987a (SN1987a)
set an upper limit for the superluminal speed different
at the order of about \(10^{-9}\) for neutrinos with energies around \(10~\rm{MeV}\) \cite{SN1987a}.

There have been many works (an incomplete list is \cite{Superluminal Works Done01,CohenG002,Dass001,Amelino-CameliaGLMRL001,PfeiferW002})
to explain the superluminal behaviors of neutrinos soon after the report of the OPERA Collaboration.
It has been well known that Einstein's special relativity refuses that
the particles move faster than the speed of light and then refuses the existence of superluminal phenomena.
Especially, it is worthwhile to note the work by Cohen and Glashow in which they pointed out that
such superluminal neutrinos would be suppressed since they lose their energies rapidly via the bremsstrahlung of electron-positron
pairs (\(\nu\longrightarrow\nu+e^{-}+e^{+}\)) \cite{CohenG002}.
Thus, if the reported superluminal behaviors of neutrinos is true, Einstein's special relativity should be amended in principle.
Minkowski spacetime is the base of Einstein's special relativity.
To account for the superluminal phenomena, one new type of spacetime structure as well as special relativity based on it is required and essential.
In such a new spacetime, the so-called superluminal speed should still hold the casuality.
This means that the speed of particles should be slower than the casual speed.

Finsler spacetime \cite{Book by Bao,Book by Shen} provides a reasonable framework to solve the superluminal problem mentioned above.
It is the generalization of the Riemann spacetime without constraint on the quadratic form of the metric,
which introduces new characters into the spacetime background.
The geometry of Lorentz invariant violation has been studied \cite{RatzelRS001},
which often points to the deformed dispersion relations derived from Finsler geometry \cite{Deformed DR,PfeiferW002}.
The casual structure has also been studied in Finsler spacetime and the possibilities of superluminal behaviors of particles
have been pointed in bimetric Finsler spacetime in reference \cite{PfeiferW001,PfeiferW002}.
It is noted that the so-called superluminal behaviors in Finsler spacetime do not contradict with causality,
and the causal speed could be faster than $1$.
Therefore, the law of causality consists with the superluminal neutrinos in such a new spacetime.
It is well known that the Lorentz group is replaced by its proper subgroup 
in very special relativity (VSR) proposed by Cohen and Glashow \cite{CohenG001} in 2006.
Soon after that, the line element of VSR was proved to be of Finslerian type \cite{VSR Finsler},
which makes VSR becomes a Finslerian special relativity.
Another example of Finslerian special relativity \cite{ChangL001} resides in Randers spacetime \cite{Randers space}.
In Randers spacetimes, there is one more extra 1-form than Minkowski spacetime, which introduces Lorentz invariant violation.
The symmetry and special relativity in Finsler spacetime with constant curvature have been studied systematically \cite{LiC001}.
Finsler geometry has been shown powerful in the studies of gravity and cosmology
to resolve the anomalies of the standard theories \cite{Finsler gravity}.

In this Letter, we propose that the spacetime background should be described by a class of Finsler spacetimes.
We study the kinematics and causal structure, and reveal the superluminal phenomena of neutrinos in such a Finsler spacetime.
The new dispersion relation is obtained.
It is found that superluminal speed of a particle is approximately linearly dependent on the energy per unit mass.
The formula is consistent with data of the current muon neutrino experiments.
Furthermore, the theoretic prediction is also consistent with the observations on electron neutrinos from SN1987a.
It is worthwhile to note that evidences of the superluminal speed difference with linear energy dependence have been presented
from best fitting for experimental data in previous works \cite{Dass001,Amelino-CameliaGLMRL001}.

In Einstein's special relativity, the line element of Minkowski spacetime is given by
\begin{equation}
\label{Special relativity}
ds=\bar{F}d\tau=\sqrt{\eta_{\mu\nu}y^{\mu}y^{\nu}}d\tau\ ,
\end{equation}
where \(\eta_{\mu\nu}=(+1,-1,-1,-1)\) is the metric tensor of Minkowski spacetime.
The over-bars denote that the objects exist in Minkowski spacetime.
It is obvious that Minkowski or more generally Riemann spacetime is a special case of Finsler spacetime.
Let the canonical 4-momentum of a particle with mass $m$ be defined as
\begin{equation}
\bar{p}_{\mu}:=m\frac{\partial \bar{F}}{\partial y^{\mu}}=\frac{my_{\mu}}{\bar{F}}\ .
\end{equation}
Then the energy and the 3-momentum could be defined as
\begin{eqnarray}
\bar{E}&:=& \sqrt{\eta^{00}\bar{p}_{0}\bar{p}_{0}}\ ,\\
\bar{P}&:=& \sqrt{-\eta^{ij}\bar{p}_{i}\bar{p}_{j}}\ .
\end{eqnarray}
The speed of a particle in Minkowski spacetime could be given by
\begin{equation}
\bar{v}:=\frac{\bar{P}}{\bar{E}}\ .
\end{equation}

%
%
%
%

The kinematics in Finsler geometry \cite{Book by Bao,Book by Shen} originates from the integrals of the form
\begin{equation}
\label{integral length}
\int^b_a F\left(x^{\mu}, y^{\nu}\right)d\tau\ ,
\end{equation}
where $x^{\mu}$ denotes position and \(y^{\mu}:=dx^{\mu}/d\tau\) denotes velocity of a particle.
Note that Greek letters run from $0$ to $3$ and Latin letters run from $1$ to $3$ in this Letter.
The Finsler structure $F\left(x, y\right)$ is defined on the slit tangent bundle $TM\setminus0$ instead of the manifold $M$.
A Finsler structure is a positive-definite function with the property of positive homogeneity of degree one \cite{Book by Bao}
\begin{equation}
\label{Finsler structure}
F(x,\lambda y)=\lambda F(x,y)\ ,
\end{equation}
for all $\lambda>0$.
Associated with a Finsler structure on the slit tangent bundle, a manifold could be viewed as a Finsler manifold.
The metric tensor in Finsler manifold is defined as an \(n\times n\) Hessian matrix of the form:
\begin{equation}
\label{metric tensor}
g_{\mu\nu}(x,y):=\frac{\partial}{\partial y^\mu}\frac{\partial}{\partial y^\nu}\left(\frac{1}{2}F^2\right)\ .
\end{equation}
The metric tensor is used to lower and raise the indices of vectors and tensors together with its inverse.

We follow the ordinary definitions of canonical energy-momentum and speed of the particles
while suppose that spacetime structure is described by the one in Finsler spacetime,
which departures mildly from the Minkowski structure.
In this Finsler spacetime, the spacetime background is described by the line element of the \((\alpha,\beta)\) type \cite{alpha beta type}
\begin{equation}
\label{Finsler Superluminal}
ds=F(y^{\mu})d\tau=\left(\alpha+A\frac{{\beta}^{3}}{{\alpha}^{2}}\right)d\tau\ ,
\end{equation}
where
\begin{eqnarray}
\alpha&=&\sqrt{\eta_{\mu\nu}y^{\mu}y^{\nu}}\ ,\\
\beta&=&b_{\mu}y^{\mu}\ .
\end{eqnarray}
Here, $b_{\mu}$ denotes the constant vector \(\left(1,0,0,0\right)\) and $A$ denotes one tiny positive dimensionless constant (\(A\ll 1\)).
This means that there is just a mild extension to Einstein's special relativity,
and it is obvious that this extension would vanish in the case that \(A=0\).
In addition, this spacetime is contained in the class of so-called locally Minkowski spacetime \cite{Book by Bao}
which is flat in Finsler geometry.
%
%
%

\begin{figure}[h]
\begin{center}
\includegraphics[width=8cm]{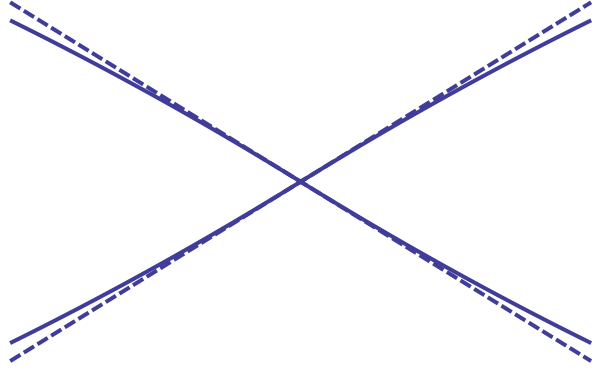}
\caption{A schematic plot for the Finslerian null structure. The dashed line denotes the null structure in the Einstein's special relativity
and the real line denotes the one in Finsler spacetime.
It is obvious that the null structure is enlarged in this Finsler spacetime.}
\label{fig1}
\end{center}
\end{figure}

First, we should study the null structure \(F=0\) in this Finsler spacetime.
The null structure would give the causal structure of the spacetime background or the dispersion relation of the massless particles.
The 4-velocity of a massless particle in Finsler spacetime could be defined as
\begin{equation}
\label{velocity in null structure}
u_{\mu}:=\frac{\partial F}{\partial y^{\mu}}\ ,
\end{equation}
with normalization \(F=0\).
The null structure could be revealed as
\begin{equation}
g^{\mu\nu}u_{\mu}u_{\nu}=F^{2}=0\ .
\end{equation}
We could rewritten this null structure to be
\begin{equation}
{\alpha}^{2}+2A\frac{{\beta}^{3}}{F}
\approx0\ ,
\end{equation}
and then renormalize this equation with $F$ as what we do in Einstein's special relativity
\begin{equation}
\eta_{\mu\nu}\frac{dx^{\mu}}{d\lambda}\frac{dx^{\nu}}{d\lambda}+2A\left(\frac{dx^{0}}{d\lambda}\right)^{3}
=0\ ,
\end{equation}
where we have neglected the terms of higher orders in $A$.
In addition, the parameter $\lambda$ denotes the corresponding affine parameter after the normalization with $F$.
This null structure is illustrated schematically in Fig.\ref{fig1}.
Then, we obtain the causal speed in this Finsler spacetime to be
\begin{equation}
v_{c}=\frac{\sqrt{-\eta_{ij}\frac{dx^{i}}{d\lambda}\frac{dx^{j}}{d\lambda}}}{\sqrt{\eta_{00}\frac{dx^{0}}{d\lambda}\frac{dx^{0}}{d\lambda}}}
\approx1+A\frac{dx^{0}}{d\lambda}\ .
\end{equation}
This null structure and causal speed in Finsler structure are enlarged in comparison to the ones in Minkowski structure
since the parameters $A$ is a positive constant.

In the following, we step forwards to the discussions on the kinematics of particles with mass in the Finsler spacetime.
The action of a free particle with mass $m$ in Finsler spacetime is given by
\begin{equation}
I=\int L d\tau=m \int F(y^{\mu}) d\tau\ .
\end{equation}
%
%
Similarly as that in Minkowski spacetime, the canonical energy-momentum in Finsler spacetime could be defined as
\begin{equation}
p_{\mu}:=m\frac{\partial F}{\partial y^{\mu}}\ ,
\end{equation}
with normalization \(F(y)=1\).
%
Then the new dispersion relations in Finsler spacetime is of the form
\begin{equation}
g^{\mu\nu}p_{\mu}p_{\nu}=m^{2}\ .
\end{equation}
It is always hold since the Finsler structures are homogeneous of order one.
Thus, we obtain the new dispersion relation with Finslerian line element (\ref{Finsler Superluminal}) as
\begin{equation}
\label{MDR}
\eta_{\mu\nu}p^{\mu}p^{\nu}+\frac{2A}{m}\left(p^{0}\right)^{3}
=m^{2}\ ,
\end{equation}
where we have neglected the terms related to $A$ in higher orders than two.

\begin{figure}[h]
\begin{center}
\includegraphics[width=10cm]{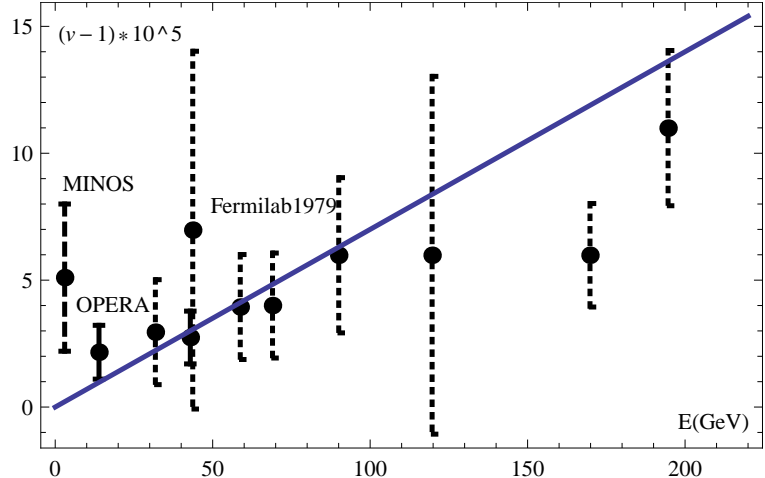}
\caption{The plot for predicted speed difference formula (\ref{difference of velocities}).
The data of OPERA \cite{OPERA01} (solid bars), MINOS \cite{MINOS01} (dashed bars)
and Fermilab1979 \cite{Fermilab197902} (dotted bars) on muon neutrino experiments
have also been revealed for comparison with this prediction.
The parameter $A$ is set to be \(7\times10^{-18}\).}
\label{fig0}
\end{center}
\end{figure}

From the dispersion relation (\ref{MDR}), we could get the speed of a particle as
\begin{equation}
\label{superluminal speed formula}
v=1-\frac{1}{2u^{2}}+Au\ ,
\end{equation}
where, for convenience, we have written the speed of a particle as function of the energy per unit mass \(u:=\frac{E}{m}\).
It is obvious that the particles could propagate faster than 1 when the constant \(A\) is positive.
Furthermore, this speed of particle is not contradictory with the causal speed
since it is slower than the causal speed by an extra positive term \(\frac{1}{2u^{2}}\).
So that the causality still holds while permits existence of superluminal neutrinos in the Finslerian special relativity.
The upper limit of muon neutrino is at the order of  $0.01~\rm{eV}$ \cite{PDG}.
We can neglect the term \(\frac{1}{2u^{2}}\) in Eq.(\ref{superluminal speed formula}) 
if the energies of neutrinos are higher than \(1~\rm{MeV}\).
From the Eq.(\ref{superluminal speed formula}), the speed difference between the speed of neutrinos and 1 is approximately
\begin{equation}
\label{difference of velocities}
\delta v=v-1\approx Au\ ,
\end{equation}
which is linearly dependent on the energies per unit mass of the particles $u$.
The prediction for the speed difference formula (\ref{difference of velocities}) is revealed in Fig.\ref{fig0}.
This predicted result could be compared with the data of
OPERA \cite{OPERA01}, MINOS \cite{MINOS01} and Fermilab1979 \cite{Fermilab197902} muon neutrino experiments.
It is clear that the Finslerian special relativity is consistent with these experiments on superluminal neutrinos
and the parameter $A$ is constrained to be at the order of \(10^{-18}\).
In addition, the superluminal speed difference is found to be at the order of $10^{-9}$ for neutrinos with energies around \(10~\rm{MeV}\),
which is also consistent with the observations on electron neutrinos from SN1987a.
Furthermore, the latest Michelson-Morley experiment \cite{MullerSTIWHSKP001} showed that Lorentz invariant violation
for photons with lower energies \(0.1~\rm{eV}\) was limited to be less than \(10^{-16}\).
This result is also consistent with the predictions of Finslerian special relativity.

Conclusions and remarks are listed as follows.
In this Letter, we have shown that Finsler spacetime permits the existence of superluminal behaviors of particles.
A Finsler structure (\ref{Finsler Superluminal}) was proposed to describe the spacetime
which mildly departures from the Minkowski one.
The null structure was studied and the causal speed was obtained.
It is found that the null structure and causal speed is enlarged in this Finsler structure.
The new dispersion relation (\ref{MDR}) was obtained in such a Finsler spacetime.
Then we investigated kinematics in this Finsler spacetime.
It was found that the Finslerian special relativity admits the superluminal phenomena
and could account for the reported superluminal behaviors of neutrinos.
In the new speed formula, there is one unique parameter $A$ which leads to Lorentz invariant violation.
The superluminal speed difference formula is linearly dependent on the energies per unit mass of particles.
We compared this formula with the data of the OPERA, MINOS and Fermilab1979 muon neutrino experiments in Fig.\ref{fig0}.
It was found that the Finslerian speed formula is consistent with the data of these neutrino experiments
and the parameter $A$ was constrained to be order of \(10^{-18}\).
Furthermore, we showed that the Finslerian special relativity is also consistent with the superluminal constraint
set by the observations on electron neutrinos from SN1987a.

\begin{acknowledgments}
This work is supported by the National Natural Science Fund of China under Grant No. 10875129 and No. 11075166.
One of us (S. Wang) thank Y. G. Jiang, M. H. Li, X. J. Bi, P. F. Yin and X. G. Wu for some useful discussions.
\end{acknowledgments}


\begin{thebibliography}{999}

\bibitem{OPERA01}T. Adam et al. [OPERA Collaboration], arXiv: 1109.4897.

\bibitem{MINOS01}P. Adamson at al. [MINOS Collaboration], Phys. Rev. D {\bf 76}, 072005 (2007).

\bibitem{Fermilab197901}J. Alspector et al., Phys. Rev. Lett. {\bf 36}, 837 (1976).

\bibitem{Fermilab197902}G. R. Kalbfleisch, N. Baggett, E. C. Fowler, and J. Alspector, Phys. Rev. Lett. {\bf 43}, 1361 (1979).

\bibitem{SN1987a}K. Hirata et al., Phys. Rev. Lett. {\bf 58}, 1490 (1987).
    R. M. Bionta et al., Phys. Rev. Lett. {\bf 58}, 1494 (1987).
    M. J. Longo, Phys. Rev. D {\bf 36}, 3276 (1987).


\bibitem{Superluminal Works Done01}
    G. Cacciapaglia, A. Deandrea and L. Panizzi, arXiv: 1109.4980v1 [hep-ph].
    F. Cardone, R. Mignani and A. Petrucci, arXiv: 1109.5289v2 [physics.gen-ph].
    K. Cahill, arXiv: 1109.5357v3 [physics.gen-ph].
    F. Tamburini and M. Laveder, arXiv: 1109.5445v5 [hep-ph].
    J. Ciborowski and J. Rembielinski, arXiv: 1109.5599v1 [hep-ex].
    F. R. Klinkhamer, arXiv: 1109.5671v3 [hep-ph].
    G. F. Giudice, S. Sibiryakov and A. Strumia, arXiv: 1109.5682v1 [hep-ph].
    G. Dvali and A. Vikman, arXiv: 1109.5685v1 [hep-ph].
    S. S. Gubser, arXiv: 1109.5687v2 [hep-th].
    R. Alicki, arXiv: 1109.5727v1 [physics.data-an].
    M. Fayngold, arXiv: 1109.5743v1 [physics.gen-ph].
    R. B. Mann and U. Sarkar, arXiv: 1109.5749v2 [hep-ph].
    A. Drago, I. Masina, G. Pagliara and R. Tripiccione, arXiv: 1109.5917v1 [hep-ph].
    M. Li and T. Wang, arXiv: 1109.5924v1 [hep-ph].
    J. Magueijo, arXiv: 1109.6055v1 [hep-ph].
    Z. Lingli and B. Q. Ma, arXiv: 1109.6097v1 [hep-ph].
    V. Pankovic, arXiv: 1109.6121v2 [physics.gen-ph].
    C. R. Contaldi, arXiv: 1109.6160v1 [hep-ph].
    V. K. Oikonomou, arXiv: 1109.6170v1 [hep-th].
    R. A. Konoplya, arXiv: 1109.6215v1 [hep-th].
    L. Iorio, arXiv: 1109.6249v1 [gr-qc].
    S. Hannestad and M. S. Sloth, arXiv: 1109.6282v1 [hep-ph].
    J. Alexandre, J. Ellis and N. E. Mavromatos, arXiv: 1109.6296v1 [hep-ph].
    A. Kehagias, arXiv: 1109.6312v1 [hep-ph].
    A. Nicolaidis, arXiv: 1109.6354v1 [hep-ph].
    Z. Lingli and B. Q. Ma, arXiv: 1109.6387v2 [hep-th].
    S. Gardner, arXiv: 1109.6520v2 [hep-ph].
    F. R. Klinkhamer and G. E. Volovik, arXiv: 1109.6624v1 [hep-ph].
    M. Matone, arXiv: 1109.6631v1 [hep-ph].
    E. Ciuffoli, J. Evslin, J. Liu and X. Zhang, arXiv: 1109.6641v1 [hep-ph].
    S. Gardner, arXiv: 1109.6520v2 [hep-ph].
    X. J. Bi, P. F. Yin, Z. H. Yu and Q. Yuan, arXiv: 1109.6667v1 [hep-ph].
    K. Whisnant, arXiv: 1109.6860v1 [hep-ph].
    P. Wang, H. Wu and H. Yang, arXiv: 1109.6930v1 [hep-ph].
    G. Cacciapaglia, A. Deandrea and L. Panizzi, arXiv: 1109.4980v1 [hep-ph].
    B. Koch, arXiv: 1109.5721v1 [hep-ph].
    R. Alicki, arXiv: 1109.5727v1 [physics.data-an].
    R. B. Mann and U. Sarkar, arXiv: 1109.5749v2 [hep-ph].
    M. Li and T. Wang, arXiv: 1109.5924v1 [hep-ph].
    G. Bhattacharyya, H. P\(\ddot{a}\)s and D. Pidt, arXiv: 1109.6183v1 [hep-ph].
    R. Garattini and G. Mandanici, arXiv: 1109.6563v1 [gr-qc].
    F. R. Klinkhamer and G. E. Volovik, arXiv: 1109.6624v2 [hep-ph].
    L. Gonzalez-Mestres, arXiv: 1109.6630v1 [physics.gen-ph].
    M. Matone, arXiv: 1109.6631v2 [hep-ph].
    E. Ciuffoli, J. Evslin, J. Liu and X. Zhang, arXiv: 1109.6641v2 [hep-ph].
    X. J. Bi, P. F. Yin, Z. H. Yu and Q. Yuan, arXiv: 1109.6667v1 [hep-ph].
    P. Wang, H. Wu and H. Yang, arXiv: 1109.6930v3 [hep-ph].
    M. M. Anber and J. F. Donoghue, arXiv: 1110.0132v1 [hep-ph].
    J. Franklin, arXiv: 1110.0234v1 [physics.gen-ph].
    G. Henri, arXiv: 1110.0239v1 [hep-ph].
    R. Cowsik, S. Nussinov and U. Sarkar, arXiv: 1110.0241v1 [hep-ph].
    R. Torrealba, arXiv: 1110.0243v1 [hep-ph].
    E. Canessa, arXiv: 1110.0245v2 [physics.gen-ph].
    S. Y. Li, arXiv: 1110.0302v1 [hep-ph].
    W. Winter, arXiv: 1110.0424v2 [hep-ph].
    J. M. Carmona and J.L. Cortes, arXiv: 1110.0430v1 [hep-ph].
    P. Wang, H. Wu and H. Yang, arXiv: 1110.0449v1 [hep-ph].
    T. Li and D. V. Nanopoulos, arXiv: 1110.0451v1 [hep-ph].
    I. Ya. Aref'eva and I. V. Volovich, arXiv: 1110.0456v1 [hep-ph].
    G. Amelino-Camelia, L. Freidel, J. Kowalski-Glikman and L. Smolin, arXiv: 1110.0521v1 [hep-ph].
    J. Knobloch, arXiv: 1110.0595v3 [hep-ex].
    B. Broda, arXiv: 1110.0644v1 [physics.data-an].
    E. N. Saridakis, arXiv: 1110.0697v1 [gr-qc].
    R. Ehrlich, arXiv: 1110.0736v4 [hep-ph].
    C. S. Unnikrishnan, arXiv: 1110.0755v1 [physics.gen-ph].
    R. Brustein and D. Semikoz, arXiv: 1110.0762v1 [hep-ph].
    L. Maccione, S. Liberati and D. M. Mattingly, arXiv: 1110.0783v1 [hep-ph].
    H. Davoudiasl and T. G. Rizzo, arXiv: 1110.0821v1 [hep-ph].
    X. Y. Wu, X. J. Liu, N. Ba, B. J. Zhang and Y. Wang, arXiv: 1110.0882v1 [physics.gen-ph].
    S. Nojiri and S. D. Odintsov, arXiv: 1110.0889v1 [hep-ph].
    I. Oda and H. Taira, arXiv: 1110.0931v1 [hep-ph].
    D. V. Naumov and V. A. Naumov, arXiv: 1110.0989v1 [hep-ph].
    D. V. Ahluwalia, S. P. Horvath and D. Schritt, arXiv: 1110.1162v1 [hep-ph].
    A. Mecozzi and M. Bellini, arXiv: 1110.1253v1 [hep-ph].
    S. S. Xue, arXiv: 1110.1317v1 [hep-ph].
    J. W. Moffat, arXiv: 1110.1330v1 [hep-ph].
    S. Tanimura, arXiv: 1110.1790v1 [hep-ph].
    A. E. Faraggi, arXiv: 1110.1857v1 [hep-ph].
    M. G. Ivanov, arXiv: 1110.1875v1 [hep-ph].
    C. Y. Zhu, H. Fan and S. P. Ding, arXiv: 1110.1943v1 [hep-ph].
    F. Henry-couannier, arXiv: 1110.2060v1 [hep-ph].
    B. Altschul, arXiv: 1110.2123v1 [hep-ph].
    F. R. Klinkhamer, arXiv: 1110.2146v1 [hep-ph].
    A. Stebbins, arXiv: 1110.2170v1 [hep-ex].
    A. Stebbins, arXiv: 1110.2170v1 [hep-ex].
    E. Capelas de Oliveira, W. A. Rodrigues Jr. and J. Vaz Jr, arXiv: 1110.2219v1 [math-ph].
    S. Sahu and B. Zhang, arXiv: 1110.2236v1 [hep-ph].
    T. R. Morris, arXiv: 1110.2463v1 [hep-ph].

\bibitem{CohenG002}A. G. Cohen and S. L. Glashow, arXiv: 1109.6562v1 [hep-ph].

\bibitem{Dass001}N. D. H. Dass, arXiv: 1110.0351v1 [hep-ph].

\bibitem{Amelino-CameliaGLMRL001}G. Amelino-Camelia, G. Gubitosi, N. Loret, F. Mercati, G. Rosati, P. Lipari, arXiv: 1109.5172v2 [hep-ph].

\bibitem{PfeiferW002}C. Pfeifer and M. N. R. Wohlfarth, arXiv: 1109.6005v1 [gr-qc].

\bibitem{Book by Bao}D. Bao, S. S. Chern, and Z. Shen, {\it An Introduction to Riemann--Finsler Geometry},
        Graduate Texts in Mathmatics {\bf 200}, Springer, New York, 2000.

\bibitem{Book by Shen}Z. Shen, {\it Lectures on Finsler Geometry}, World Scientific, Singapore, 2001.

\bibitem{RatzelRS001}D. Ratzel, S. Rivera, F. P. Schuller, Phys. Rev. D {\bf 83}, 044047 (2011); arXiv: 1010.1369v2 [hep-th].

\bibitem{Deformed DR}G. Amelino-Camelia, J. R. Ellis, N. E. Mavromatos and D. V. Nanopoulos, Int. J. Mod. Phys. A {\bf 12}, 607 (1997); arXiv: hep-th/9605211v1.
    R. Gambini, J. Pullin, Phys. Rev. D {\bf 59}, 124021 (1999); arXiv: gr-qc/9809038v1.
    F. Girelli, S. Liberati and L. Sindoni, Phys. Rev. D {\bf 75}, 064015 (2007); arXiv: gr-qc/0611024v3.
    F. P. Schuller, C. Witte and M. N. R. Wohlfarth, Annals Phys. {\bf 325}, 1853 (2010); arXiv: 0908.1016v1 [hep-th].

\bibitem{PfeiferW001}C. Pfeifer and M. N. R. Wohlfarth, Phys. Rev. D {\bf 84} 044039 (2011); arXiv: 1104.1079 [gr-qc].

\bibitem{CohenG001}A. G. Cohen, S. L. Glashow, Phys. Rev. Lett. {\bf 97}, 021601 (2006); arXiv: hep-ph/0601236.

\bibitem{VSR Finsler}G. W. Gibbons, J. Gomis, C. N. Pope, Phys. Rev. D {\bf 76}, 081701 (2007).
    G. Y. Bogoslovsky, arXiv: 0706.2621.
    H. F. Goenner and G. Y. Bogoslovsky, Gen. Rel. Grav. {\bf 31}, 1383 (1999); arXiv: gr-qc/9701067.
    A. P. Kouretsis, M. Stathakopouslos, and P. C. Stavrinos, Phys. Rev. D {\bf 79}, 104011 (2009); arXiv: 0810.3267.





\bibitem{ChangL001}Z. Chang and X. Li, Chinese Phys. C {\bf 33}, 626 (2009); arXiv: 0809.4762v1 [gr-qc].

\bibitem{Randers space}G. Randers, Phys, Rev. {\bf 59}, 195(1941).

\bibitem{LiC001}X. Li and Z. Chang, arXiv: 1010.2020v2 [gr-qc].

\bibitem{Finsler gravity}
    Y. Takano, Lett. Nuovo Cimento {\bf 10}, 747 (1974).
    S. Ikeda, Ann. der Phys. {\bf 44}, 558 (1987).
    R. Tavakol, and N. van den Bergh, Phys. Lett. A {\bf 112}, 23 (1985).
    G. Yu. Bogoslovsky, Phys. Part. Nucl. {\bf 24}, 354 (1993).
    Z. Chang and X. Li, Phys.Lett.B {\bf 668}, 453 (2008).
    M. Milgrom, Astrophys. J. {\bf 270}, 365 (1983).
    Z. Chang and X. Li, Phys. Lett. B {\bf 676}, 173 (2009).
    X. Li, Z. Chang and M. H. Li, arXiv: 1001.0066.
    Z. Chang, M. H. Li and X. Li, arXiv: 1009.1509.
    X. Li and Z. Chang, Phys. Lett. B {\bf 692}, 1 (2010).
    J. D. Anderson {\it et al.}, Phys. Rev. Lett. {\bf 81}, 2858 (1998).
    J. D. Anderson {\it et al.}, Phys. Rev. Lett. {\bf 65}, 082004 (2002).
    J. D. Anderson {\it et al.}, Mod. Phys. Lett. A {\bf 17}, 875 (2002).
    X. Li and Z. Chang, Phys. Rev. D {\bf 82}, 124009 (2010).
    D. Clowe, S. W. Randall, and M. Markevitch, http://flamingos.astro.ufl.edu/1e0657/index.html; Nucl. Phys. B, Proc. suppl. {\bf 173}, 28 (2007).
    X. Li and Z. Chang, arXiv: 0911.1890.
    Z. K. Silagadze, Acta Phys. Polon. B {\bf42}, 1199-1206 (2011); arXiv: 1007.4632.
    Z. Chang, S. Wang and X. Li, arXiv: 1106.2726.
    X. Li and Z. Chang, arXiv: 1108.3443.

\bibitem{alpha beta type}Z. Shen, {\it Some perspectives in Finsler geometry}, MSRI Publication Series. Cambridge: Cambridge university press, 2004.

\bibitem{PDG}C. Amsler {\it et al.}, ``Review of Particle Physics'', Phys. Lett. B {\bf 667}, 1 (2008); doi:10.1016/j.physletb.2008.07.018.

\bibitem{MullerSTIWHSKP001}H. M$\ddot{u}$ller, P. L. Stanwix, M. E. Tobar, E. Ivanov, P. Wolf, S. Herrmann, A. Senger, E. Kovalchuk and A. Peters, Phys. Rev. Lett. {\bf 99}, 050401 (2007).


\end{thebibliography}
\end{document}